\documentclass[twocolumn,pre,showpacs,amsmath,amssymb]{revtex4}

\usepackage{hyperref}
\usepackage{graphicx}
\usepackage{bm}
\usepackage{times}

\newcommand{\Teff}{T_\mathrm{eff}}
\newcommand{\TTST}{T_{\mathrm{eff},\mathsf{TST}}}
\newcommand{\Tludo}{T^\ast}
\newcommand{\Tfast}{T_\mathrm{fast}}
\newcommand{\Tslow}{T_\mathrm{slow}}
\newcommand{\rTST}{r_\mathsf{TST}}
\newcommand{\ave}[1]{\langle #1 \rangle}

\begin{document}
\title{Driven activation versus thermal activation}

\author{Patrick Ilg\footnote{Present address: ETH Z\"urich,
Polymer Physics, HCI H541, CH-8093 Z\"urich, Switzerland}}
\author{Jean-Louis Barrat}
\affiliation{Universit\'e de Lyon; Univ. Lyon I,  Laboratoire de
Physique de la Mati\`ere Condens\'ee et des Nanostructures; CNRS,
UMR 5586, 43 Bvd. du 11 Nov. 1918, 69622 Villeurbanne Cedex, France}
\date{\today}

\begin{abstract}

Activated dynamics in a glassy system undergoing steady shear
deformation is studied by numerical simulations. Our results show
that the external driving force has a strong influence on the
barrier crossing rate, even though the reaction coordinate is only
weakly coupled to the nonequilibrium system. This "driven
activation" can be quantified by introducing in the Arrhenius
expression an effective temperature, which is close to  the one
determined from the fluctuation-dissipation relation. This
conclusion is supported by analytical results for a simplified model
system.
\end{abstract}


\maketitle

Activated rate theory  is ubiquitous in the description  and
understanding of dynamical processes in condensed matter, physical
chemistry or materials science. The basic problem, known as the
"barrier crossing" or "Kramers problem", is that of a single
degree of freedom, coupled to a heat bath, and moving in  a double
well potential. The "barrier crossing rate" is defined as the
average time taken by the system to switch from a potential well
to the other, under the influence of thermal noise. In general,
the single degree of freedom, often called "reaction coordinate",
is coupled to a complex, fluctuating environment. The "thermal
noise" is a schematic description of the interaction with this
environment.

This approach has been applied to a wealth of different problems.
We can for example mention diffusion in solids, in which case the
reaction coordinate is an atomic position, and the noise is
associated with thermal vibrations. In isomerization reactions,
the reaction coordinate is an internal coordinate of the molecule,
coupled to a liquid solvent. In nucleation theory, the internal
coordinate describes a collective fluctuation of an order
parameter, and the "barrier" is interpreted as a {\em free
energy}, rather than energy, barrier. Other examples involve the
Eyring theory of plasticity in solids, in which the activated
process is associated with a local strain change.

The analysis of the barrier crossing problem is often associated
with the names of Eyring, who proposed the so called "transition
state approximation" \cite{Eyring}, and of Kramers, who made the
first complete analysis of the problem in the limits of low and
high friction \cite{Kramers}. Since then, many refinements of the
theory have been studied and are reviewed in reference
\cite{Hanggi}. In all cases, it turns out that an essential factor
in the reaction rate, which to a large extent governs the
variation with temperature $T$, is the Arrhenius contribution:
\begin{equation}\label{arrhenius}
    r(T) \sim \exp(-\Delta E /k_B T)
\end{equation}
where $\Delta E$ is the energy barrier to overcome. The
exponential variation of the Arrhenius factor (\ref{arrhenius}) is,
in fact, the hallmark of activated processes.

As discussed above, activated processes are often invoked in the
description of the dynamical response of condensed matter systems.
As such, they will typically take place under {\em nonequilibrium}
conditions. The deviation from equilibrium can be weak,
e.g.~during the flow of a Newtonian liquid, in which case the
applicability of equation (\ref{arrhenius}) is straightforward. In
other  cases, however, the same equation is applied to systems
that are strongly out of equilibrium, in the sense that their
response to an external driving force is strongly nonlinear, or
that their phase space distribution is very different from the
equilibrium, Gibbs-Boltzmann distribution.

A prototypical example of such a strongly nonequilibrium situation
is the flow of a glassy system. Such a flow can be induced only by
stresses larger than the yield stress
(see e.g.~\cite{Robbins_Yield} for the effect of strain and temperature
in glassy solids).
In the absence of flow, the
relaxation is very slow, and the  system is out of equilibrium
and non-stationary \cite{Ludo_Science05,Poon_Science02}.
The flow produces a nonequilibrium steady
state \cite{BBK,BB,Barrat_PRL2002}, with a typical relaxation time that is fixed
internally by the applied stress or the strain rate. This
situation has attracted a considerable amount of theoretical and
experimental  interest, in two different contexts. The first one
is the rheology of "soft glasses" (emulsions, pastes, colloidal
glasses, foams). The second one is the plastic deformation of bulk
metallic glasses. In both cases, approaches have been proposed
that introduce a "noise temperature" \cite{SGR} or "disorder
temperature" \cite{Langer04,Langer05}. In \cite{SGR}, this noise
temperature replaces the actual temperature in equation
(\ref{arrhenius}). In such models, the effective temperature is
introduced in a somewhat empirical manner.

Another concept of effective temperature, rooted in statistical
mechanics ideas, was introduced in \cite{CKP,Kurchan05}, based on
the "fluctuation-dissipation ratio". At equilibrium, the
fluctuation-dissipation theorem states that the ratio between
integrated response and correlation functions  (FDR) is equal  to
the temperature. Cugliandolo \textit{et al.}~\cite{CKP} showed how
this concept could be extended to out-of-equilibrium system, by
defining the effective temperature from the FDR, which now differs
from the thermal bath temperature. It was proposed that a
thermometer probing a nonequilibrium system on long time scales
would actually be sensitive to this effective temperature, and this
result was checked numerically on simple models
\cite{BB,Barrat_PRL2002,Liu02}. Experimental evidence supporting
this definition of an effective temperature has been found e.g.~in
\cite{Makse06,Ocio02}.

In this contribution, we explore the influence of an external
driving force on the rate of a simple activated process. Our primary
objective is here to check how the external drive, and the "noise"
it generates, can influence the dynamics of an internal degree of
freedom, which is not directly coupled to the driving force. A very
standard way of quantifying the results is to use the Arrhenius
representation, which provides an operational  way of introducing an
"activation temperature", that can be compared to other calculations
of effective temperatures in nonequilibrium systems.

\begin{figure}[h]
  \begin{minipage}{4cm}
  \includegraphics[width=4cm]{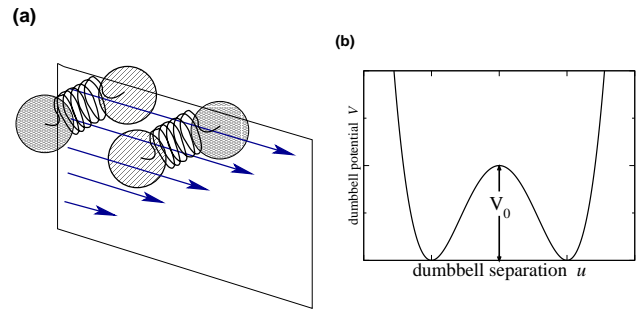}
  \end{minipage}\hspace{0.2cm}
  \begin{minipage}{4cm}
  \includegraphics[width=4cm]{ilgbarrat_fig1b.eps}
  \end{minipage}
  \caption{(a) Schematic representation of a dumbbell particle in a system
  undergoing shear flow with fixed orientation perpendicular to the
  shear plane. Also shown is an isomerization reaction. The
  magnitude of the separation between the two centers of force is
  considerably exaggerated in this schematic representation.
  (b) Intramolecular potential (characteristic of the nonlinear
  "spring" shown in panel (a) ) between the two centers of force that define
  the dumbbell.}
  \label{fig1}
\end{figure}

Our approach  involves the simulation of the classical Kob-Andersen
"binary Lennard-Jones" model undergoing shear flow, similar to the
one used in ref.~\cite{BB}. In order to probe activated dynamics,
one appealing possibility would be to identify and study the
activated events that actually give rise to the flow at low
temperature, in the spirit of \cite{SGR}. This approach, however, is
difficult and could yield ambiguous results, as the flow is self
consistently coupled to these events. We therefore make use of the
flexibility of numerical modeling to devise a very simple "activated
degree of freedom" that has only a weak coupling to the existing
flow in our system. This is achieved by replacing each particle of
the minority species $\mathbf{r}_j^B$ by a peanut shaped "dumbbell"
with coordinates $\mathbf{r}_j^B\pm (u_j/2)\mathbf{e}^z$, with fixed
orientation along $\mathbf{e}^z$, the direction perpendicular to the
shear plane. Each center of force in the dumbbell carries half of
the particle interaction, and the separation between the two centers
of force $u$ is small enough that the perturbation of the
surrounding fluid can be neglected. The important feature of the
model is the fact that the two centers of force are related through
an internal "reaction coordinate" $u$, which evolves in a bistable
intramolecular potential $V(u)=(V_0/u_0^4)(u^2-u_0^2)^2$, where
$u_0=0.1$ (in Lennard-Jones units) is the equilibrium dumbbell
separation (see fig.~\ref{fig1}b). Each dumbbell is therefore a
simple "two-state" system which can undergo, under the influence of
the interactions with the surrounding fluid, an "isomerization
reaction". This reaction corresponds to exchanging the positions of
the two centers of force (see fig.~\ref{fig1}).

This "isomerization" will be the focus of our study. Its rate can
be studied as a function of the imposed barrier height, of the
external temperature $T$ and on the driving force, which is here
quantified by the shear rate $\dot{\gamma}$. We have chosen to
work under conditions for $T$ and $\dot{\gamma}$ that have been
well characterized previously \cite{BB} ($T=0.3$ and
$\dot{\gamma}=10^{-3}$, in Lennard-Jones units) and to concentrate
on the influence of barrier height $\Delta E = V_0$. At this
temperature, the system would not undergo structural relaxation on
the time scales that can be achieved using computer simulation.
Under the influence of the external drive, a relaxation on
 a time scale $\tau_\alpha \simeq 100$ is observed. This time
 scale is very well separated from microscopic, vibrational time
 scale, so that our system is a practical realization of the
theoretical concepts described in ref.~\cite{CKP}.

Determination of reaction rates is a notoriously difficult
challenge for numerical simulations, as the activated events
typically take place on much larger time scales than the short
time vibrations of the intramolecular bonds. A number of
sophisticated methods \cite{Dellago} have  been developed to
bypass this intrinsic difficulty,  either from biased simulations,
or by making use directly of the rate formula \ref{arrhenius}.
Unfortunately, such methods always assume that the system is close
to thermal equilibrium, and are therefore inapplicable in our
case.
The forward flux method recently proposed in \cite{FFS} is applicable
to nonequilibrium systems. However, only a single reaction coordinate
per system can be treated with this method, which is impractical
for the present situation.
As a result, we have to use "brute force" simulations to
obtain reaction rates from the study of individual trajectories,
which seriously limits the range of barrier heights that can be
considered.


The Sllod equations of motion
appropriate for a fluid undergoing simple shear
were integrated with a leapfrog algorithm using a
time step of $\Delta t=5\times 10^{-4}$ in reduced Lennard-Jones units
\cite{BB}.
For the dumbbell particles, the leapfrog algorithm is
applied to the center-of-mass and relative positions and momenta.
Lees-Edwards periodic boundary conditions \cite{BB} are employed
in order to minimize effects due to the finite system size.
Constant temperature conditions are ensured by rescaling the
velocity components in the neutral direction of all particles at
each time step.

The reaction rate $r$ is determined from the number correlation
function $C(t)$ by
\begin{equation} \label{correl}
C(t) = \frac{\ave{\delta n(t)\delta n(0)}}{\ave{\delta n(0)^2}}
\approx  \exp{[-r t/\ave{n}]}
\end{equation}
where $\delta n(t)=n(t)-\ave{n}$ and $n(t)$ equals one if
$u(t)>u_B$ and zero else \cite{Chandler_isomerize}.
The systems studied contain $N=2048$ particles,
$410$ of which are dumbbells.
Eq.~(\ref{correl}) is evaluated from an ensemble average over
$20-40$ independent systems.
The results are found to
be independent of the exact location of the dividing surface $u_B$ in
the vicinity of the barrier maximum $u_B=0$. The fast initial
decay of $C(t)$ is well described by transition state theory.
Escape rates are extracted from fits to Eq.~(\ref{correl}) for
intermediate times $5\leq t \leq 10$. We verified that very
similar results are found within a broad range $1\leq t \leq 30$,
before the correlation function finally decays to zero, in full
agreement with theoretical expectation \cite{Chandler_isomerize}.
For relatively low barrier heights $V_0/T\lesssim 3$,
$C(t)$ decays more rapidly, so that
we extracted rates for shorter times, $0.5\leq t\leq 1$.

In the following, we present results for the reaction
rate as a function of $V_0$ based on the study of the decay in the
number-number correlation function \cite{Chandler_isomerize}.
We adopt common
practice by giving all temperature and energy values in terms of
the depth of the Lennard-Jones potential $\epsilon$.

\begin{figure}[h]
 \includegraphics[width=9cm]{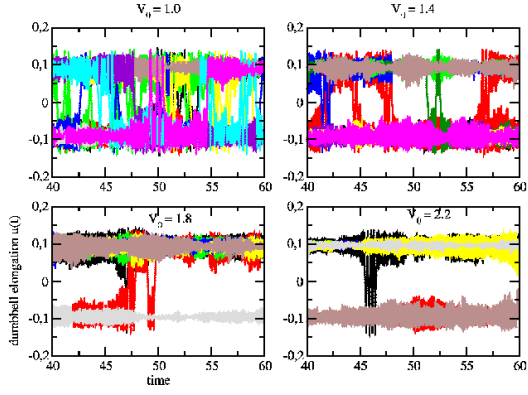}\\
  \caption{(Color online) Trajectories of the internal dumbbell coordinate for
  different barrier heights. The thermal bath temperature is  $T=0.3$ in
reduced Lennard-Jones units, and the shear rate is
$\dot{\gamma}=10^{-3}$.}
  \label{fig2}
\end{figure}

Figure \ref{fig2} illustrates the difficulty of the approach, by
showing the trajectories of selected dumbbells for different
values of the barrier at $T=0.3$.
For $V_0/kT=1$,
barrier crossings are so common that describing them trough
classical rate theory is problematic. For $V_0/kT=10$, the crossings
become very unlikely, so that the determination of the rate
becomes difficult. This leaves us with typically two
decades in terms of variation of the reaction rate.

\begin{figure}[h]
  \includegraphics[width=7cm]{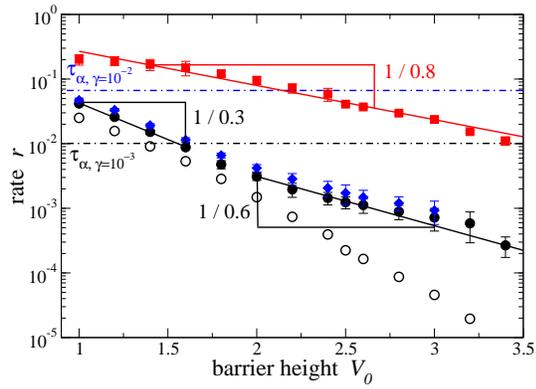}\\
  \caption{(Color online) Reaction rate as a function of barrier height, for fixed
  temperature and shear rate. Full squares: results for $T=0.8$ (red) at
  equilibrium.
  $T=0.3$ (black and blue),
  and different shear rates (full diamonds and circles
  correspond to $\dot{\gamma}=10^{-2}$ and $\dot{\gamma}=10^{-3}$,
  respectively). Open circles represent $r_\mathrm{ext}$,
  an extrapolation of the
  high temperature results to $T=0.3$ as explained in the text.}
  \label{fig3}
\end{figure}

The corresponding reaction rates, determined from the correlation
function of the dumbbell internal coordinate, are shown in figure
\ref{fig3}. 
At $T=0.8$, $\dot{\gamma}= 10^{-3}$, the rates obey the equilibrium 
Arrhenius law (\ref{arrhenius}),
showing that under these conditions the drive is only a weak perturbation
to the system.
We now concentrate on the rates obtained at $T=0.3$
and $\dot{\gamma}= 10^{-3}$. The reaction rates are clearly
influenced by the external driving imposed to the system. To show
this, we use the rates obtained
at a rather high
temperature, $r_\mathrm{eq}(T=0.8)$, to extrapolate to $T=0.3$.
The equilibrium extrapolation $r_\mathrm{ext}$ is achieved using
the Arrhenius formula, i.e. $r_\mathrm{ext}(T=0.3)=
r_\mathrm{eq}(T=0.8) \times \exp(+V_0/0.8-V_0/0.3)$. Clearly, the
extrapolated rates are significantly lower than those actually
observed under shear, except at low barrier heights (high rates)
were the two estimates almost coincide. The difference between the
extrapolated rates and the measured ones is an indicator of the
inadequacy of the standard Arrhenius formula, using the thermal
bath temperature, in the driven system.

In spite of the limited range of accessible rates, it is clear from
figure \ref{fig3} that the rates in a glassy system  under shear do
not obey Arrhenius behavior of the form $\exp(-V_0/T)$ over the
whole range of barrier heights under study. While this law is
relatively well obeyed at low barriers and large crossing rates, it
would significantly underestimate the rate for high barriers.
Instead, at  high barrier rates, the crossing rate is considerably
increased. If an attempt is made to fit the results to an "effective
Arrhenius factor", a value of $\Teff\simeq 0.6$ is obtained.

Under the same conditions, a completely different determination
of the effective temperature \cite{BB}, based on the
fluctuation-dissipation approach mentioned above,
yields $\Tludo\simeq 0.65$.
This is in good agreement with the present fit to an
Arrhenius law.
The determination of $\Teff$ based on reaction rates is of
limited accuracy, such that we cannot exclude that $\Teff$ and $\Tludo$
actually differ slightly or that $\Teff$ is slightly dependent on
barrier height. A more precise determination of $\Teff$ would require
larger barrier heights, which is computationally quite demanding.
Note, that $V_0=3$ corresponds for $T=0.3$ already to the
rather high barrier height of $V_0/kT=10$.

In figure \ref{fig3}, we also display the results obtained for the
rates at a slightly higher value of the shear,
$\dot{\gamma}=10^{-2}$. The separation of time scales between
relaxation time and microscopic times is less marked than for the
low shear rates ($\tau_\alpha\simeq 10$ in this case). It appears
that the increase in shear  induces a change in the prefactor for
the rates, rather than in the barrier height dependence. This is
consistent with the relatively weak influence of shear rate on
effective temperature reported earlier \cite{BB}.

It is interesting to discuss the time scale at which the crossover
between the two Arrhenius laws, characterized either by the bath
temperature or an effective temperature, takes place. A natural
guess would be to associate this crossover with a value of the
rate that corresponds to the inverse of the $\alpha$ relaxation
time. The general idea is, that fluctuations taking place on
longer time scales will be associated with a higher temperature
\cite{CKP,BB}. In figure \ref{fig3} we see that this guess
overestimates the crossover rate  by a factor of 5 in the case of
$\dot{\gamma}=10^{-3}$. It is not clear at this point, whether
this difference is significant or  reflects merely some
arbitrariness in the definition of relaxation times.

The simulation results presented above suggest that the activated
dynamics is governed by an elevated temperature $\Teff \simeq 0.6 >
T$. This temperature is consistent with the effective temperature
$\Tludo=0.65$ found in extensive simulation studies on the
fluctuation-dissipation relation in this system \cite{BB}. In order
to investigate the relation between $\Teff$ and $\Tludo$ and to
rationalize our simulation results, we study to following toy model
proposed in \cite{IlgBarrat06}.

Consider a particle of mass $m$ at position $x$ moving in an
external potential $V(x)$ under the influence of two thermal baths.
One bath, associated with the fast degrees of freedom, is kept at
temperature $\Tfast$ and exerts an instantaneous friction force of
strength $\Gamma_0$. The second bath, which mimics the slow degrees
of freedom is held at temperature $\Tslow$ and is described by the
retarded friction coefficient (memory kernel) $\Gamma(t)$. The
equations of motion read $\dot{x}=v$,
\begin{equation} \label{toy}
m\dot{v} = - V'(x) - \int_0^t\!ds\, \Gamma(t-s)v(s) - \Gamma_0 v(t) + \xi(t) + \eta(t)
\end{equation}
The fast bath is modeled as Gaussian white noise with
$\ave{\eta(t)}=0$, $\ave{\eta(t)\eta(s)}=2\Tfast\delta(t-s)$, whereas the
random force due to the slow bath is described by
$\ave{\xi(t)}=0$, $\ave{\xi(t)\xi(s)}=2\Tslow\Gamma(t-s)$.
We use an exponentially decaying memory kernel
$\Gamma(t)=\alpha^{-1}e^{-t/(\alpha\gamma)}$ for which
the non-Markovian dynamics (\ref{toy}) can equivalently be rewritten as Markovian
dynamics in an extended set of variables \cite{Barrat_nonmarkov90}.

Exact solutions of the model (\ref{toy}) for harmonic potentials $V$
are presented in \cite{IlgBarrat06}. For barrier crossing problems
with double-well potentials $V$, no analytical solutions to
(\ref{toy}) are known. We therefore extend the widely used
transition state approximation to the present model after adiabatic
elimination of the fast degrees of freedom. The resulting expression
for the rate $\rTST$ is rather lengthy and will be presented
elsewhere together with the (straightforward) procedure. For the
double-well potential $V(x)$ considered above, the dependence of the
rate $\rTST$ on the barrier height $V_0$ is again dominated by the
Arrhenius factor, however with an effective temperature
$\TTST=\Tfast w/[w+4(\Tfast-\Tslow)]$, where $w=\Tslow + \alpha V''
\Tfast$ and $V''=8V_0/u_0^2$. Thus, if the slow and fast bath are
both kept at the same temperature, $\Tslow=\Tfast=T$, one recovers
the usual Kramers result with $\TTST=T$. If, however, $\Tslow >
\Tfast$, the escape rate is enhanced due to $\TTST > \Tfast$. Due to
the interplay between fast and slow dynamics in the barrier
crossing, the effective temperature is in general intermediate
between the temperature of the slow and the fast bath. These
predictions are in agreement with the simulation results presented
above. Furthermore, estimating the coefficient $\alpha\approx 0.01$
from the inverse high frequency shear modulus for the Lennard-Jones
system \cite{Barrat_nonmarkov90}, the predicted effective
temperature is $\TTST\approx 0.45$. In view of the simplicity of the
model and the uncertainty in $\alpha$, the order of magnitude
agreement with the observed $\Teff$ is reasonable.

In conclusion, we have shown that  activated processes out of
equilibrium are influenced by an external driving,  even if the
corresponding degree of freedom is weakly coupled to the drive.
Qualitatively, this increase is the essential result from our
simulations. From a more quantitative point of view, the analysis of
the Arrhenius plot allows one to define operationally an effective
activation temperature. The link  of this activation temperature to
other definitions of effective temperature, and the time scale for
the crossover from "thermal activation" to "driven activation" will
have to be explored further. However, the results are consistent
with  a general picture involving a degree of freedom coupled to two
different heat baths, one associated with short time vibrations and
one associated with shear induced fluctuations, taking place on
longer time scales and described by a higher temperature
\cite{IlgBarrat06}.

This "driven activation" (as opposed to "thermal" activation)
could have interesting consequences for characterizing the
effective temperature of nonequilibrium systems, by providing a
"thermometer" based on activated processes. It can also be of
importance within the theory of plasticity of amorphous materials,
by providing a self-consistent description of the "noise" that
induces local plastic events, within a classical statistical
mechanics description involving a noise temperature
\cite{Langer04,Caroli}.
Inserting an effective temperature in Eyring's rate theory of plasticity,
T.~Haxton and A.~J.~Liu were recently able to account for the flow curves of
simple glassy systems at low temperatures \cite{Liu_private}.

\bibliography{ilgbarrat}

\pagebreak

\bf{Figures}

\bigskip

\bigskip

\bigskip

\end{document}